\documentclass[reqno,12pt]{amsart}
\usepackage[left=3cm, top=3.2cm, bottom=3.2cm, right=3cm]{geometry}
\usepackage{amssymb}
\usepackage{amsxtra}
\usepackage{amsmath}
\usepackage{amsfonts}
\usepackage{mathrsfs}

\numberwithin{equation}{section} \allowdisplaybreaks
\newtheorem{thm}{Theorem}[section]
\newtheorem*{Mthm}{Main Theorem}

\newcommand{\eqa}{\begin{eqnarray}}
\newcommand{\eeqa}{\end{eqnarray}}
\newcommand{\beq}{\begin{equation}}
\newcommand{\eeq}{\end{equation}}
\newcommand{\nn}{\nonumber}
\newcommand{\p}{\partial}

\def \dsum{\displaystyle\sum}

\begin{document}

\title[] {Spectral Curve of the Halphen Operator}

\author[]{Andrey E. Mironov  and Dafeng Zuo}

\address {Mironov, Sobolev Institute of Mathematics, Novosibirsk, Russia,
 and  Novosibirsk State University}

\email{mironov@math.nsc.ru}

\address{Zuo, School of Mathematical Science,
University of Science and Technology of China,
 Hefei 230026,
P.R.China and Wu Wen-Tsun Key Laboratory of Mathematics, USTC,
Chinese Academy of Sciences}

\email{dfzuo@ustc.edu.cn}



\date{\today \quad \mbox{To appear in {\bf Proceedings of the Edinburgh Mathematical Society}}}
\begin{abstract}
The Halphen operator is a third-order operator of the form
$$
 L_3=\partial_x^3-g(g+2)\wp(x)\partial_x-\frac{1}{2}g(g+2)\wp'(x),
$$
where $g\ne 2\,\mbox{mod(3)}$, the Weierstrass $\wp$-function satisfies the equation
$$
 (\wp'(x))^2=4\wp^3(x)-g_2\wp(x)-g_3.
$$
In the equianharmonic case, i.e., $g_2=0$ the Halphen operator commutes
with some ordinary differential operator $L_n$ of order
$n\ne 0\,\mbox{mod(3)}.$
In this paper we find the spectral curve of the pair $L_3,L_n$.

\end{abstract}

\maketitle

\section{Introduction}
The commutativity condition $L_nL_m=L_mL_n$ for two ordinary differential
operators
$$
L_n=\partial_x^n+\sum_{j=0}^{n-2}u_{j}(x)\partial_x^{j},\quad L_m=\partial_x^m+\sum_{j=0}^{m-1}v_{j}(x)\partial_x^{j},\quad
 v_{m-1}=const
$$
is equivalent to a nonlinear system of differential equations
on their coefficients. Let us recall the first results on this system obtained
in the beginning of the last century. Wallenberg \cite{W} found operators in the simplest cases $n=2, m=3,$ and $n=2,m=5$.
Schur \cite{S} proved that
if $L_nL_m=L_mL_n$ and $L_nL_k=L_kL_n$, where $L_k$ is a differential operator
of order $k$, then $L_mL_k=L_kL_m$. So, there is a maximal commutative
ring containing $L_n$ and $L_m$.
 Burchnall and Chaundy \cite{BC} showed that if $L_n$ and $L_m$ commute,
 then there is a polynomial
$R(z,w)$ such that $R(L_n,L_m)=0$. The polynomial $R$ defines the
{\it spectral curve}
$$\Gamma=\{(z,w)\in{\mathbb C}^2: R(z,w)=0\}.$$
The spectral curve parameterizes common eigenvalues of $L_n$
and $L_m$:
$$
 L_n\psi=z\psi, \quad L_m\psi=w\psi, \quad P=(z,w)\in\Gamma.
$$
The dimension $l={\rm dim}\{L_n\psi=z\psi, L_m\psi=w\psi\}$ for
$P=(z,w)\in\Gamma$ in general position is called the {\it rank} of the pair
$L_n,L_m$. The rank is a common divisor of orders $n$ and $m$.
In the case of rank one, the function $\psi$ (the {\it Baker--Akhiezer function})
and coefficients of the operators can be expressed
through theta-function of the Jacobi variety of $\Gamma$ with the help of
Krichever
construction \cite{K1} (for the case $n=2$ see also \cite{Its}).
In the very special cases when the spectral curve covers an elliptic
curve, the coefficients of the operators can be expressed through the
Weierstrass  $\wp$-function.
 Famous examples of such operators are the Lam\'e
 operator
$$
 L_2=-\partial_x^2+g(g+1)\wp(x),
$$
the Weierstrass $\wp$-function satisfies the equation
$$
 (\wp'(x))^2=4\wp^3(x)-g_2\wp(x)-g_3,
$$
 and the Halphen operator \cite{H}
$$
 L_3=\partial_x^3-g(g+2)\wp(x)\partial_x-\frac{1}{2}g(g+2)\wp'(x),
$$
in the {\it equianharmonic} case, i.e., $g_2=0$, where $g\ne 2\,\mbox{mod(3)}$.
Lam\'e, Halphen and other operators with elliptic
coefficients have been extensively studied by Picard \cite{Pi},
Hermite and by many others, please see for instance
\cite{WW}--\cite{GV} and references therein.

The Lam\'e operator
commutes with an operator of order $2g+1$, the spectral curve is a
hyperelliptic curve of genus $g$. Grosset and Veselov \cite{GV}
showed that coefficients of the spectral curve equation could be
expressed through the
elliptic Bernoulli polynomials.

When $g_2\ne 0$ and $g\geq 7$ there is no meromorphic solution
$\psi$ to the fundamental system $L_3\psi=z\psi$, $z\in{\mathbb C}$ of
the Halphen operator (see (\cite{U})).
According to the Krichever construction \cite{K1}, eigenfunctions
of rank one commuting operators are
meromorphic functions for all values of the spectral parameter $z$.
So, the Halphen operator commutes with an operator
$L_n, n\ne 0\,\mbox{mod(3)}$ for $g\geq 7$ only if $g_2=0$.
We thus restrict our study to
the case $g_2=0$.
The Halphen operator commutes with an operator of order $3r+1+\epsilon$ for
$\epsilon=0$ or $1$  and the spectral curve is a trigonal curve of genus
$g=3r+\epsilon$.
This spectral curve for small $g$ was found in \cite{DGU1999}--\cite{EEP}.
The main result of this paper is to find the spectral curve
for all possible $g$ in the equianharmonic case.  More precisely,

\begin{Mthm}In the equianharmonic case $g_2=0$, the spectral curve
of the Halphen operator  is given by the equation
   \beq w^3=F_g(z), \quad {\rm deg} F_g(z)=g+1,\label{u1.1}\eeq
{\rm (1)} when $g=6M$ or $g=6M+3$,
\beq F_g(z)= \dfrac{1}{1296}\Big(36 z^2 A_0^2+g_3 \big(2 g A_0+g^2 A_0+12 g_3 A_1\big)^2\Big)
\Big(36 z^2 A_0-\nn\eeq
\beq
 -g_3g(g+2)(7 g^2+14 g-24) A_0 - 24g_3^2 A_1(7 g^2+14 g-180) -2880 g_3^3 A_2\Big),\label{u1.2}
\eeq
$A_j$ is given in \eqref{WZ3.25};

{\rm (2)} when $g=6M-2$ or $g=6M+1$, the function $F_g(z)$ is defined by the formula \eqref{MZ3.8}.
 \end{Mthm}

This paper is organized as follows. In section 2, we recall the definition of the Baker--Akhiezer function and
how to factorize skew-symmetric
operators of third order. In section 3, using results of the section 2 we prove the main theorem and list some examples.

\section{Factorization of rank one commuting operators}

\subsection{Baker--Akhiezer function}
Let $\Gamma$ be a Riemann surface, $q$ a fixed point on $\Gamma$, $k^{-1}$ a local parameter near $q$ and
$\gamma_1+\dots+\gamma_g$ a nonspecial  divisor on $\Gamma$, where $g$ is the genus of $\Gamma$.
The Baker--Akhiezer function $\psi(x,P)$, $P\in\Gamma$ is a unique function which satisfies the following two
conditions (see \cite{K1}):

1. In the neighborhood of $q$ it has the form
$
 \psi=e^{xk}\left(1+\frac{\xi_1(x)}{k}+\frac{\xi_2(x)}{k^2}+\dots\right).\nn
$

2. On $\Gamma\backslash\{q\}$, the function $\psi$ is meromorphic with $g$ simple poles $\gamma_1,\dots,\gamma_g$.

\noindent For the meromorphic function $f(P),\ P\in\Gamma$ with the unique pole in $q$, there is a unique differential operator
$L_f$ such that
$
 L_f\psi=f(P)\psi.
$
For such kinds of functions $f(P)$ and $g(P)$, the corresponding operators $L_f$ and $L_g$ commute.

Let us consider the first order operator vanishing on $\psi$
$$
 (\partial_x-\chi(x,P))\psi(x,P)=0,\qquad \chi(x,P)=\frac{\psi_x(x,P)}{\psi(x,P)}.
$$
The function $\chi$ is a meromorphic function on $\Gamma$ with simple poles $P_1(x),\dots,P_g(x)$ and $q$. In the
neighborhood of $q$ it has the form
$
 \chi(x,P)=k+O(1).
$
Let $k-\gamma_i(x)$ be a local parameter near $P_i(x)$, then in the neighborhood of $P_i(x)$
$$
 \chi(x,P)=\frac{-\gamma_i'(x)}{k-\gamma_i(x)}+O(1).
$$
The operator $\partial_x-\chi$ is a right common divisor of
$L_f-f(P)$ and $L_g-g(P)$, i.e.
$$
 L_f-f(P)=\tilde{L}_f(\partial_x-\chi),\qquad L_g-g(P)=\tilde{L}_g(\partial_x-\chi).
$$

\subsection{Schr\"odinger operator}
Before discussing the third-order operators we briefly discuss the Schr\"o\-dinger operators and demonstrate
our ideas on simpler case of the Lam\'e operator.

S.P. Novikov proved \cite{N} that if the smooth real periodic
Schr\"odinger operator $L_2=-\p_x^2+u(x)$
commutes with an operator $L_{2g+1}$
 then it has finite number of gaps in
its spectrum (the inverse statement was proved by B.A. Dubrovin \cite{D}).
The spectral curve related to the commuting pair $L_2$ and $L_{2g+1}$ is a hyperelliptic curve
$$
 w^2=z^{2g+1}+c_{2g}z^{2g}+\dots+c_0.
$$
The operator $L_2-z$ has the following factorization (see for example \cite{DMN1976})
$$
 L_2-z=-(\p_x+i\chi_0)(\p_x-i\chi_0),
\qquad \chi_0=-\frac{iQ_x}{2Q}+\frac{w}{Q},\qquad \chi=i\chi_0,
$$
where $Q=z^g+\alpha_{g-1}(x)z^{g-1}+\dots+\alpha_0(x).$
The function $Q$ satisfies the equation
\beq 4w^2=4(z-u(x))Q^2-Q_x^2+Q Q_{xx}.\label{WZ2.3}\eeq
Furthermore, taking a differentiation on $x$, and so
\beq Q_{xxx}+4(z-u(x))Q_x-2u_x(x)Q=0. \label{WZ2.4}\eeq
It turns out that for the Lam\'e operator the function $Q$ has the form (see \cite{M5})
\beq
 Q=A_g\wp^g(x)+A_{g-1}(z)\wp^{g-1}(x)+\dots+A_0(z),\nn
\eeq
where $A_g$ is a constant, $A_k=0$ for $k>g,$ and
$$
 A_s=\frac{(s+1)(8A_{s+1}z-A_{s+2}g_2(s+2)(2s+3)-
 2A_{s+3}g_3(s+2)(s+3))}{4(2s+1)(g^2+g-s(s+1))}.
$$

\begin{thm}{\rm (\cite{M5})} The spectral curve of the Lam\'e operator is given by the equation
\beq
 w^2=\frac{1}{4}(4A_0^2z-A_0(4A_2g_3+A_1g_2)+A_1^2g_3).\nn
\eeq
\end{thm}
Self-adjoint commuting rank two operators $L_4=(\partial_x^2+V(x))^2+W(x)$, $L_{4g+2}$ were studied in \cite{M6}.
An analogues of the equations (\ref{WZ2.3}) and (\ref{WZ2.4}) were obtained. With the help of this equations the first examples
of rank two operators for $g>1$ were constructed in \cite{M5}, \cite{M6}.

\subsection{Skew-symmetric operator of third order}
Let us consider a skew-symmetric operator
$L_3=\p_x^3+f(x)\p_x+\frac{1}{2} f_x(x).$
We assume that $L_3$ commutes with an operator $L_N$ of order
$N=3r+1+\epsilon$, where $\epsilon$ is 0 or 1. The spectral curve
related to the pair $L_3,$  $L_N$ is given by
$w^3=H_g(z)w+F_g(z),$
where $g=N-1$ and  ${\rm deg}F_g(z)=N$ and $0\leq{\rm deg}H_g(z)\leq 2r+1.$
We factorize $L_3-z$ by
\beq L_3-z=(\p_x^2+\chi\p_x+\eta)(\p_x-\chi), \quad \eta=f(x)+\chi^2+2\chi_x.\label{WZ3.3}\nn\eeq
The function $\chi=\chi(x,z)$ satisfies the equation
$ \chi_{xx}+3\chi\chi_x+f(x)\chi+\chi^3+\frac{1}{2} f_x(x)=z.$
In \cite{DGU1999} it was proved that
$
 \chi=\frac{(S+\frac{1}{2}Q_x)w+Q_2}{Q w +Q_1},
$
where $Q$ and $S$ are polynomial in $z$ with
coefficients depending on $x$, if $g=3r$, then ${\rm deg} Q(z)=r-1,\ {\rm deg} S(z)=r,$ if $g=3r+1$, then
${\rm deg}Q(z)=r,\ {\rm deg}S(z)=r-1,$ and
\beq  Q_1= S^2+\frac{1}{3}Q^2f+\frac{1}{3}Q Q_{xx}-\frac{1}{4}Q_x^2,\nn\eeq
\beq Q_2=(S+\frac{1}{2} Q_x)(S_x-\frac{1}{6} Q_{xx})
+\Big(\frac{1}{3}Q_x-S_{xx}-\frac{2}{3} f S+\frac{1}{6} Q_{xxx}
\Big)Q+\Big(z+\frac{1}{6}f_x\Big) Q^2.\nn\eeq
Functions $Q$ and $S$ satisfy the following equations
\beq 3zQ_x=2S_{xxx}+2fS_x+Sf_x, \label{WZ3.9}\eeq
\beq 8 f^2 Q_x+36 z S_x+9 Q_x f_{xx}+15 f_x Q_{xx}+2 Q f_{xxx}+2 f (4 Q f_x+5 Q_{xxx})+2\partial_x^5Q=0. \label{WZ3.10} \eeq
 The function $\chi$
has $g$ simple poles $P_1(x),\dots,P_g(x)\in\Gamma$ and a simple
pole at infinity.
By using $Q$ and $S$, one could obtain

\begin{thm}{\rm (\cite{DGU1999})} The spectral curve $\Gamma$
of the pair $L_3, L_N$ is given by the equation
 $w^3=H_g(z)w+F_g(z),$
where
\eqa
H_g(z)&=&\frac{1}{12} \Big(4 f^2 Q^2+12 S_x^2+2 Q^2 f_{xx}+
Q_{xx}^2+(10 Q Q_{xx}-12 S^2-5 Q_x^2)f\nn\\&& -\, {24 S S_{xx}-2
Q_x Q_{xxx}+} {(36 z S+5 f_x Q_x+2
\partial_x^4Q)Q\Big)}\label{WZ3.11}\eeqa
and
 \eqa
F_g(z)&=& {\frac{1}{432} } {\Big(432 z S^3-63 f_x Q_x^3+}
{4 Q^3 (108 z^2+8 f^3-3 f_x^2+6 f f_{xx})}\nn \\
&-&{42 f Q_x^2 Q_{xx}+144 S_x^2 Q_{xx}-4 Q_{xx}^3-}
{432 Q_xS_x S_{xx}+12 Q_x Q_{xx} Q_{xxx}}\nn\\
&-&{36 S (3 z Q_x^2+16 f Q_xS_x-}
{4 Q_{xx} S_{xx}+4 S_x Q_{xxx})-18 Q_x^2 \partial_x^4Q}\nn\\
&+& {36 S^2 (7 f_x Q_x+10 f Q_{xx}+2 \partial_x^4Q)+}
{6 Q (12 f^2 (4 S^2-Q_x^2)-}
24 S f_x S_x      \nn\\
&+& 72 z Q_x S_x+
{12 S^2 f_{xx}-3 Q_x^2 f_{xx}}+ {16 f_x Q_x Q_{xx}+72 S_{xx}^2-2 Q_{xxx}^2}
\nn\\
&+& {2 f (12 S_x^2+7 Q_{xx}^2+48 S S_{xx}-}
{4 Q_x Q_{xxx})+4 Q_{xx} \partial_x^4Q)}+12 Q^2 (10 f^2 Q_{xx}\nn\\
&+&2 f_{xx} Q_{xx}-
{72 z S_{xx}-2 f_x Q_{xxx)}+}
f (-36 z S+3 f_x Q_x+2 \partial_x^4Q))\Big).\label{WZ3.12}
 \eeqa
\end{thm}

 \section{ The Spectral Curve of the Halphen Operator}
 In this section we assume that
$ f(x)=-g(g+2)\wp(x), \quad g
\ne 2~\hbox{mod(3)} \in \mathbb{N},$
where $\wp(x)$ is the
equianharmonic Weierstrass function.
The corresponding operator $L_3$ is the Halphen operator.

 \subsection{The proof of the main theorem}
 \begin{proof}
Substituting $f(x)=-g(g+2)\wp(x)$ into \eqref{WZ3.9}, \eqref{WZ3.10} and eliminating $Q(x,z)$,
one can get the equation on $S(x,z)$
\beq
m_0 S+m_1 S_x+m_2 S_{xx}+\dsum_{j=3}^8 m_j S^{(j)}=0,\label{MZ3.1}
\eeq
where
\eqa && m_0=16 (g-3) (g-1) g (2+g) (3+g) (5+g) \wp(x)^2
{(g_3+4 \wp(x)^3) \wp'(x)};\nn\\
&& m_1=9 g_3 (g (2+g) (5 g^2-16+10 g) g_3+12 z^2)+
(2 g (2+g) (1080-918 g-347 g^2\nn\\&&\quad+112 g^3+28 g^4) g_3+
108 z^2) \wp(x)^3+ 128 g (g-3) (g-1) (2+g) (3+g) (5+g) \wp(x)^6;\nn\\
&& m_2=((9 g (2+g) (16-10 g-5 g^2) g_3-108 z^2) \nn\\&&\qquad
 +16 (g-1) g (2+g) (3+g) (2 g+g^2-30)\wp(x)^3){\wp(x)\wp'(x)};\nn\\
 && m_3=12 g (2+g) (57-22 g-11 g^2) g_3 \wp(x)^2+
 {312 g (g-1) (2+g) (3+g) \wp(x)^5};\nn\\
 && m_4=12 g (-4 (2+g) g_3+3 (g-1) (2+g) (3+g) \wp(x)^3)
{\wp'(x)};\nn\\
&& m_5=48 g (2+g) \wp(x) (g_3+\wp(x)^3);\quad m_6=24 g (2+g) \wp(x)^2 \wp'(x);
\nn\\
&& m_7=4 (g_3+10 \wp(x)^3);\quad
m_8=-4 \wp(x) \wp'(x).\nn
\eeqa
(1) We assume that $g=6M$ or $g=6M+3$.
In this case, our main observation is that for the Halphen operator
the function $S$ has the form
\beq
 S(x,z)=\dsum_{r=0}^{M} zA_{r}(z) \wp(x)^{3r}.\label{MZ3.2}
\eeq
In order to find coefficients $A_r(z)$, let us substitute \eqref{MZ3.2} into \eqref{MZ3.1} and obtain
\beq \dsum_{r=0}^{M} A_{r}(z)
\Big(a_{r} +b_{r}\wp(x)^{3}+c_{r}\wp(x)^{6}+d_{r}\wp(x)^{9}\Big)\wp(x)^{3r}=0,\label{MZ3.3}\eeq
where
\eqa
a_r&=&108 {g_3}^3 (r-2) (r-1) r (8-3 r) (5-3 r) (4-3 r) (3 r-2) (3 r-1);\nn\\
b_r&=&216{g_3}^2 (r-1)r(3 r-5) (3 r-2) (3 r-1)(4g-30+2g^2+87r\nn\\
 &-&6gr-3g^2r-81r^2+54r^3);\nn\\
c_r&=&9r(3r-2)g_3\Big(4g^2-48g +28g^3+7g^4+264gr+84g^2r-48g^3r \nn\\&-&
12g^4r+3024r^2-4104gr^2-1620g^2r^2+432g^3r^2+108g^4r^2+1728gr^3\nn\\&+&864g^2r^3+45360r^4
-10368gr^4-5184 g^2r^4+46656r^6\Big)+324r(3r-2)z^2;\nn\\
d_r&=&8(2+3r)(3+g+3r)(1+6r)(2+g+6r)(5+g\nn\\&+&6r)(6r-g)(g-3-6r)(6r+1-g).\nn
\eeqa
Thus, it follows from \eqref{MZ3.3} that a recursive formula of $A_{r}(z)$ is given by
\beq A_{r}(z)=\frac{1}{d_{r}}\Big(c_{r+1}A_{r+1}(z)+b_{r+2}A_{r+2}(z)
+a_{r+3}A_{r+3}(z) \Big), \label{WZ3.25}\eeq
where $ A_{r}(z)=a_{r}=b_r=c_r=d_r=0$ when $r<0$ or $r>M$.
If $g=6M$, we choose $A_M(z)=\dfrac{const}{z}$; and if $g=6M+3$, we choose $A_M(z)=const$.

 Furthermore, from \eqref{WZ3.9} and \eqref{WZ3.10}, we have
\eqa Q(x,z)&=&\dsum_{r=0}^M\frac{6r(2-2 g-g^2+18 r+36 r^2)-g (2+g)}{3 (3r+1)}
 A_{r}(z) \wp(x)^{1+3 r}\nn \\
&+& \dsum_{r=0}^M \frac{2 g_3 r(9 r^2-1) }{(3 r+1)}A_{r}(z)\wp(x)^{3 r-2}.\nn\eeqa
Let $x_0$ be a zero of $\wp(x)$, then
\beq
 (\wp'(x_0))^2=g_3,\quad \wp''(x_0)=0,\quad \wp^{(3)}(x_0)=0,\quad \wp^{(4)}(x_0)=0.\label{MZ3.6}
\eeq
Substituting $S(x,z)$ and $Q(x,z)$ at $x=x_0$ into \eqref{WZ3.11} and \eqref{WZ3.12} and
using the above identities in \eqref{MZ3.6}, we obtain
$H_g(z)=0$ and the equation \eqref{u1.1} with $F_g(z)$ given in \eqref{u1.2}
which is the first part of our main theorem.

(2) We assume that $g=6M-2$ or $g=6M+1$. In this case,
our main observation is the function $S$ being of the form
$S(x,z)=\dsum_{r=1}^{M} zB_{r}(z) \wp(x)^{3r-1},$
where $B_r=0$ for $r<1$ and $r>M$. If $g=6M-2$, we choose $B_M(z)=\dfrac{const}{z}$;
and if $g=6M+1$, we choose $B_M(z)=const$.
By analogy with the above process, we could obtain
\beq B_{r}(z)=\frac{1}{d_{r}}\Big(c_{r+1}B_{r+1}(z)+b_{r+2}B_{r+2}(z)
+a_{r+3}B_{r+3}(z) \Big), \label{MZ3.7}\eeq
and $H_g(z)=0$ and
\beq
F_g(z)=\frac{1}{62208 g^3 (g-6+4 g^2+g^3)^3}
 \Big((25 (384g-1728+172 g^2-20g^3-5g^4) g_3B_1\nn \eeq
\beq + 108z^2B_1-1440 (6 g+3 g^2-140) g_3^2 B_2-80640 g_3^3 B_3)(-64 g^2 (g+4 g^2+g^3-6)^2 g_3^2(324 B_1^2 z^2
\nn\eeq
\beq+25 g_3 ((2 g+g^2-24) B_1+
24 g_3 B_2)^2)+16 g (6-g-4 g^2-g^3) g_3 (-108 z^2 B_1+5 g (2+g) g_3 \nn\eeq
\beq((-24+2 g+g^2) B_2+24 g_3 B_2))
 (25 (384 g-1728+172 g^2-20 g^3-5 g^4) g_3 B_2+108 z^2 B_1\nn\eeq
\beq-1440 (6 g+3 g^2-140) g_3^2 B_2-80640 g_3^3 B_3)-(36 z^2-4 g^2 g_3-4 g^3 g_3-g^4 g_3) (25 (1728-384 g\nn\eeq
\beq-172 g^2+20 g^3+5 g^4) g_3 B_1+108 z^2 B_1-1440 (6 g+3 g^2-140)
g_3^2 B_2\nn \eeq
\beq-80640 g_3^3 B_3)^2)\Big).\label{MZ3.8}\eeq
We thereby complete the proof of our main theorem.
\end{proof}



\subsection{Examples}
To end up this section we give some examples to illustrate the above
construction of the spectral curve $\Gamma$: $w^3=F_g(z)$ and auxiliary functions $S(x,z)$ and $Q(x,z)$:

\noindent 1. $g=1:$ $S=0,\quad Q=1,\quad F_1(z)=z^2.$

\noindent 2. $g=3:$ $ S=z,\quad Q=-5 \wp(x),\quad F_3(z)=z^4-\frac{55 g_3z^2}{2}-\frac{3375 g_3^2}{16}.$

\noindent 3. $g=4:$ $S=-56 \wp(x)^2,\quad Q=z,\quad F_4(z)=z^5-208 g_3 z^3+12544 g_3^2 z.$

\noindent 4. $g=6:$  $S=(z^2-880g_3)+3520 \wp(x)^3,\quad Q=-16 z \wp(x),$
$$F_6(z)=z^7-2992g_3z^5+2972416g_3^2z^2-1003622400 g_3^3 z .$$

\noindent 5. $g=7:$ $S=-420 z \wp(x)^2,\quad Q=z^2-\frac{8775 g_3}{4}+9100 \wp(x)^3,$
$$
 F_7(z)=z^8-8151 g_3 z^6+\frac{175837875 g_3^2 }{8}z^4
 -\frac{309670034375 g_3^3 }{16}z^2-\frac{109044078609375 g_3^4}{256}.
$$
\noindent 6. $g=9:$ $S=z^3-\frac{53823g_3z}{4}+39424z\wp(x)^3,$
$$Q=\frac{1145375}{4}g_3\wp(x)-33 z^2\wp(x)-1172864\wp(x)^4,$$
$$
 F_{9}(z)=z^{10}-\frac{167739 g_3}{4}z^8+\frac{4760523141 g_3^2}{8}z^6-
 \frac{95260137283003 g_3^3}{32}z^4
$$
$$
 +\frac{428576521043796741 g_3^4 }{256}z^2
+\frac{236605250703471890625 g_3^5}{1024}.\
$$

\noindent 7. $g=10:$ $S=22131200 g_3 \wp(x)^2-1560 z^2 \wp(x)^2-88524800 \wp(x)^5,$
$$Q=z^3+83200 z \wp(x)^3-25920 g_3 z,$$
$$F_{10}(z)=z^{11}-83600 g_3 z^9+2409504000 g_3^2 z^7-26083604480000 g_3^3 z^5$$
$$+63684041113600000 g_3^4 z^3-50781428593459200000 g_3^5 z.$$





 \noindent{\bf Acknowledgments.} We are grateful to the anonymous referee's suggestions
 for improvement.  The research of A. Mironov is partially
supported by grant 12-01-00124-a from the Russian Foundation for Basic Research;
program of the Presidium of RAS ''Fundamental Problems of Nonlinear Dynamics in Mathematical and Physical Sciences'';
 and a grant from Dmitri Zimin's
``Dynasty" foundation.  The research of D.\,Zuo is partially
supported by NCET-13-0550 and NSFC (11271345, 11371338) and the Fundamental Research
Funds for the Central Universities. Part of this work was done while the first author stayed in
School of Mathematical Science, University of Science and Technology
of China. He is grateful to the university for kind hospitality.


\end{document}